\documentclass[preprint,amsmath,amssymb,showpacs]{revtex4}
\usepackage{graphicx}
\usepackage{dcolumn}
\usepackage{bm}
\newcommand{\ek}{\epsilon_k}
\begin{document}
\newcommand{\etatwo}{\frac{\eta}{2}}
\newcommand{\etaone}{\eta}
\renewcommand{\etatwo}{\eta}
\renewcommand{\etaone}{2\eta}
\newcommand{\etafour}{ {\frac{\etatwo}{2}}}

\title{Discretized Thermal Green's Functions}
\author{M. Granath, A. Sabashvili, H. U. R. Strand, and S. \"Ostlund}
\affiliation{Department of Physics, University of Gothenburg,
SE-41296 Gothenburg, Sweden}

\date{\today}

\begin{abstract}
We present a spectral weight conserving formalism for Fermionic
thermal Green's functions that are discretized in imaginary time
$\tau$ and thus periodic in imaginary (``Matsubara'') frequency
$i\omega_n$.  The formalism requires a generalization of the Dyson
equation $G(G_0,\Sigma)$ and the Baym-Kadanoff-Luttinger-Ward
functional for the free energy $\beta\Omega=\Gamma(G)$.  A conformal
transformation is used to analytically continue the periodized
Matsubara Green's function to real frequencies in a way
that conserves the discontinuity at $t=0$ of the corresponding
real-time Green's function.  This
allows numerical Green's function calculations of very high precision
and it appears to give a well controlled convergent approximation
in the $\tau$ discretization.  The formalism is tested
on dynamical mean field theory calculations of the paramagnetic
Hubbard model.
\end{abstract}


\maketitle
The analytic properties of finite temperature Green's functions is
a cornerstone of the theory of quantum many body
theory.\cite{ref:abrikosov,N_O} The Green's functions can be represented
either in continuous complex time  or by discrete values on an
infinite set of Matsubara frequencies. The physical retarded Green's
function and the corresponding spectral function is given by the analytic continuation of this discrete
function of Matsubara frequencies to a continuous function on the
real frequency axis.  Although it is extremely elegant the method
poses a numerical challenge.  
It was early recognized that the Pad\'e series  could be used fit
data from a finite number of Matsubara frequencies.\cite{vidberg}
This scheme is numerically quite ill conditioned and much effort
has been devoted to resolving this difficulty. For quantum Monte
Carlo methods the Maximum Entropy method \cite{Jarrell} that directly
computes the spectral function as a probabilility distribution is instead
widely used.


We develop a method to work with a Fermion Green's function defined exactly
only at $ N $ equally spaced points  $ \tau_j =  \frac{\beta}{N} j$ in imaginary time.
In this space of dimension $ N $ a discrete solution is computed 
numerically exactly. A conformal transformation is used to
construct an exact analytic continuation using a rational function. Since
$ N $ does not have to be large to yield stable results, calculations 
can be done using very high precision.  This resolves many of the 
difficulties others have had with an ill conditioned Pad\'e series.\cite{numexact} 

Our approach assures that several basic conditions are satisfied.
In addition to the obvious demand that the limit $ N \rightarrow \infty $ 
yields a proper continuum limit, we obtain three additional
properties valid for all $ N $ that are not present in previous
approaches and which give a well controlled and rapidly convergent
result as $ N $ is increased. For {\em all } values of $ N $ we
find that (a) the Green's function obeys exactly a Luttinger-Ward
variational principle (b) the free energy is {\em exact } for
noninteracting particles (c) a numerically exact analytic continuation
exists that reproduces the data on the imaginary frequency axis and has
the proper discontinuity and analytic structure at $ t = 0 $.

Let us consider Fermions described by a
Hamiltonian $H=H_0+V$ with $H_0=\sum_k\ek c^\dagger_kc_k$,
and where $c^\dagger_k$ represents Fermion creation 
operators  of a state $|k\rangle$ and where $k$ may be momentum and spin.  
The chemical potential $ \mu$ is absorbed in $\ek$.  The interaction is
$V=\frac{1}{2}\sum_{k,k',q,q'}V_{kk'qq'}c^\dagger_kc^\dagger_{k'}c_{q'}c_q$.
We consider the one-particle Green's function
$G_k(\tau,\tau')=-\langle T(c_k(\tau)c_k^\dagger(\tau') \rangle =-\frac{1}{Z}Tr\{e^{-\beta
H} T(e^{\tau H}c_ke^{-\tau H}e^{\tau' H}c_k^\dagger e^{-\tau' H})\}$ where
$Tr$ is the sum over a complete set of states, $T$ is time ordering,
$\beta$ is inverse temperature, and $Z$ is the partition function
$Z=Tr\{ e^{-\beta H} \}$.  We assume
$G$ is diagonal in $k$ and there are no anomalous ``superconducting'' terms.
The function $ G_k( \tau, \tau') $ is defined on $-\beta<\tau-\tau'<\beta$ and
obeys $G_k(\beta-\tau)=-G_k(-\tau )$. This antiperiodicity condition allows
$ G_{k}(\tau)  $ to be transformed to Matsubara frequencies
$i\omega_n=\frac{2\pi \, i }{\beta}(n+\frac{1}{2})$ for integer $n$.

The expansion of $ G_k(\tau) $ in powers of $V$ can be expressed
as sums of connected diagrams consisting of the vertex $V$ at a
time $\tau$ and the non-interacting Green's function
$G_{0,k}(\tau-\tau')=-\langle T(c_k(\tau)c^\dagger(\tau')\rangle_0$,
where internal times are integrated over as $\int_0^\beta d\tau$
and an $n$'th order diagram has a prefactor 
$\frac{(-1)^{n+n_l}}{2^n n!}$ with $n_l$ the number of Fermion loops.\cite{N_O} 
In this work we discretize $ \tau $ and replace integrals over continuous
time $0<\tau<\beta$ with sums over  
$\tau_j $, by carefully defining a discretized Green's function
and self energy.

We start by defining $ \eta = \frac{\beta}{2N}  $
and $\Omega_N=\frac{\pi}{\eta}$ and discretize the non-interacting Green's function 
$G_{0,k}(\tau)=e^{- \tau\ek}[n_f(\ek)\Theta(-\tau+0^+)+(n_f(\ek)-1)\theta(\tau-0^+)]$,
with $n_f(\ek)=1/(e^{\beta\ek}+1)$, to times $\tau_j = \etaone j  $ . The value at
$\tau_0=0$  is defined as the average of the two limits so that
$G_{0,k}(\tau_0)=n_f(\ek)-\frac{1}{2}$. 
We can expand $G_{0,k}(\tau_j)=\frac{1}{\beta}\sum_{n=0}^{N-1}e^{-i\omega_n\tau_j}G_{0,k}(i\omega_n)$ and
the ``Matsubara transform''  $ G(i \omega_n) = {\cal F} ( G( \{ \tau_j \} ) ) $ is 
\begin{equation}
G_{0,k}(i\omega_n)=  2\eta \sum_{j=0}^{N-1}e^{i\omega_n\tau_j}G_{0,k}(\tau_j)=
	\eta\coth\, \eta(i\omega_n-\ek)\,. \label{eq:coth}
\end{equation}  
For compactness we will usually drop the index $k$ and frequency
$i\omega_n$ when it is clear from the context.  We see that $G_0$
is periodic\cite{note1} under $i\omega_n\rightarrow i\omega_n+i\Omega_N$ and
in the limit $N\rightarrow\infty$ ($\eta\rightarrow 0$) the continuum
expression $G_0 =1/(i\omega_n-\ek)$ is appropriately recovered. We
also define the additional Green's functions 
$G_0^{\pm}=G_0 \pm \eta$ in analogy to the 
$ \tau = 0^{\mp} $ limit of $ G_{0}(\tau) $.

Let us now define the periodized full Green's function $G_k(i\omega_n)$ and the
self-energy $ \Sigma_k(i\omega_n) $ through the two expressions
\begin{equation}
G_k(i\omega_n)=\eta\coth\eta(i\omega_n-\ek-\Sigma_k(i\omega_n))
\label{eq:new_Dyson}
\end{equation}
and 
$\Sigma_k(i\omega_n)=\frac{\delta\Phi}{\delta G_k(i\omega_n)}\,.$
The object $\Phi$ is the functional\cite{LW,Baym_Kadanoff}
defined as the sum of linked closed skeleton diagrams of $G_k(i\omega_n)$, except
for the 1st order diagrams where we use
$G_k^+(i\omega_n)=G_k(i\omega_n)+\eta$. The self energy is thus given
by the amputated skeleton diagrams and the resulting set of equations
using Eq. \ref{eq:new_Dyson}. 

Equation \ref{eq:new_Dyson} is a generalization of the standard Dyson equation
$G^{-1}_k(i\omega_n)=i\omega_n-\ek-\Sigma_k(i\omega_n)$, and
reduces to the latter as $\eta\rightarrow 0$. It also preserves the
property that a constant, $k$ and $\omega_n$ independent, self energy
acts a chemical potential. 
\newcommand{\Sigmabar}{{\bar{\Sigma}}}
Defining 
$ \Sigmabar  \equiv  \frac{1}{\eta} \tanh \left( \eta \Sigma_k ( i \omega_n ) \right) $
which reduces to $ \Sigma_k ( i \omega_n ) $ in the limit $ \eta \rightarrow 0 $.
Eq. \ref{eq:new_Dyson} can also be rewritten in the more suggestive form 
\begin{equation} \label{eq:newnewdyson}
G^{-1} =\frac{G_0^{-1}- \Sigmabar}{1-{\eta}^2 G_0^{-1}\Sigmabar} \,,
\end{equation}
which again reduces to the ordinary Dyson equation for $ \eta \rightarrow 0 $. 

Consider now the free energy $ \Omega = - \frac{1}{\beta} \ln Z $. As 
shown by Luttinger and Ward \cite{LW}, for the standard formalism, the free energy can be expressed as 
$ \beta \Omega=\Gamma\equiv\Phi( \{G\}) - Tr \Sigma G + Tr \log( - G ) $, where 
$ Tr \equiv \sum_{k,\omega_n } $. The expression $\Gamma=\Gamma(\{G\})$ also provides a
variational formulation where $\Gamma=\beta\Omega$ corresponds to a
stationary point $ \delta\Gamma/ \delta G = 0 $ that yields
the Dyson equation $ G^{-1} = G_0^{-1} - \Sigma $.
We now describe how to generalize this variational formulation to be
consistent with Eq. \ref{eq:newnewdyson} and demands (a)-(b) stated
in the introduction. 

We define the following generalization
\begin{equation} \label{eq:discretefree}
\Gamma = \Phi( \{ G \}) -
Tr (G^{+}\Sigma) +Tr \log\left(-G^{-}/(2\eta) \right)  
\end{equation}
where $G^\pm=G\pm\eta$. The expression $\Gamma$ reduces to the standard expression 
in the limit $\eta\rightarrow 0$ and 
the stationarity condition $\delta\Gamma/\delta G =0$ gives the 
generalized Dyson equation Eq. \ref{eq:newnewdyson}.  In addition, it 
can be shown that in the noninteracting limit when $ \Sigma= \Phi = 0 $ 
the free energy in Eq.\ref{eq:discretefree} is {\em exact } for all values of $ N $
through the expression
$\sum_{k,n}\ln\frac{N}{\beta}(-G^-_{0,k}(i\omega_n))=-\sum_k\ln(1+e^{-\beta\ek})=\beta\Omega_0.$
The functional $ \Gamma( \{ G \}) $ can also be shown to give a 
value for the total particle number which is consistent with 
the formalism i.e.  
$n=-\frac{d\Omega}{d\mu}=\frac{1}{\beta} Tr \,G^{+} $.
Baym and Kadanoff \cite{Baym_Kadanoff}
noted that conserving approximations can be
found by including only a finite number of diagrams in 
the Luttinger-Ward function. This construction
can now be used for the discretized Green's functions to include
only a subset of diagrams in $ \Phi $.

Let us now explore the analytic structure of the periodized Green's
function. Starting with the spectral representation in terms of a
complete set of eigenstates $H|n\rangle=E_n|n\rangle$, 
for $\tau>0$, 
$G_k(\tau)=-\frac{1}{Z}\sum_{m,n}e^{-\beta E_n}e^{\tau
  (E_n-E_m)}|\langle m|c_k^{\dagger}|n\rangle|^2\,$ and
using Eq. \ref{eq:coth}, we can compute  $G$ as
\begin{equation}
\label{GofA}
G_k(i\omega_n) = \int_{-\infty}^{\infty}\frac{d\omega}{2\pi} \, 
	A(k,\omega)\eta\coth \eta(i\omega_n-\omega)\,,
\end{equation} 
where the spectral function is given by the conventional expression 
$A(k,\omega)=\frac{1}{Z}\sum_{m,n}|\langle m|c_k^{\dagger}|n\rangle|^2e^{-\beta
  E_n}(1+e^{-\beta\omega})2\pi\delta(E_m-E_n-\omega) $.
Eq. \ref{GofA} gives the analytic continuation
to $z=\omega+i Im(z)$ by letting $i\omega_n\rightarrow z$ 
and shows that $G_k(z)$ is analytic except where $Im(z)$ is
an integer multiple of $\Omega_N=\frac{\pi}{\eta}$.  Using 
$  \eta  Im  \coth{ \eta ( \omega_0  + i 0^+  - \omega ) } = - \pi \delta( \omega_0 - \omega ) $ 
and defining $G(z=\omega+i0^+)\equiv G^R_k(\omega)$ and
$G(z=\omega+i0^-)\equiv G^A_k(\omega)$, we formally recover the standard expression
$A(k,\omega) = \pm2 Im G^{A/R}_k(\omega)$.
Inserting into Eq. \ref{GofA} gives 
the generalized Kramers-Kronig relation 
\begin{equation} 
Re G^R_k(\omega')=
-\frac{\eta}{\pi}P\int_{-\infty}^{\infty}d\omega\frac{Im G^R_k(\omega)}{\tanh{\eta}(\omega'-\omega)}\,.  
\end{equation}

\begin{figure} 
\includegraphics[width=17.5cm]{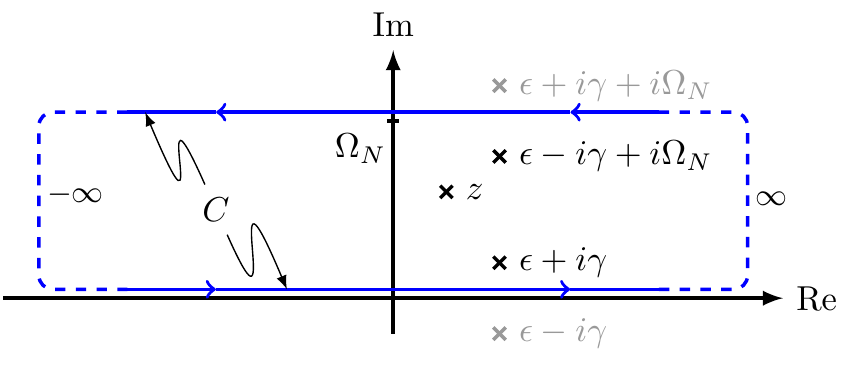}
\caption{\label{contour} (Color online) Contour for the integration of Eq.
\ref{Gfirst}} \end{figure}

We will now show how to use a conformal transformation to make an analytic continuation 
of the periodized Green's function to a rational function with simple poles.  Consider a
spectral function of the form
\begin{equation} L_{\epsilon,\gamma}(\omega)=
\frac{i {\etatwo}}{\sinh{\etatwo}(\omega-\epsilon+i\gamma)}- \frac{i
{\etatwo}}{\sinh{\etatwo}(\omega-\epsilon-i\gamma)} .
\end{equation}
which reduces to the Lorentzian
$\frac{2\gamma}{(\omega-\epsilon)^2+\gamma^2}$ in the limit
$\eta\rightarrow 0$. Because of the antiperiodicity and to ensure
positive spectral weight we can assume $0 \le  \gamma<\Omega_N/2$. 
We can evaluate the integral in Eq. \ref{GofA}  by a closed contour
containing three poles, as shown
in Figure \ref{contour}.  The result, for $0<Im z<\Omega_N$, is 
\begin{eqnarray} \label{Gfirst}
G(z)=\frac{1}{2}\oint_C\frac{dz'}{2\pi}L_{\epsilon,\gamma}(z'){\eta}\coth{\eta}(z-z')=\\
\frac{\eta}{2}\left( \coth\frac{\eta}{2}(z-\epsilon+i\gamma)+\tanh \frac{\eta}{2} (z-\epsilon-i\gamma) \right) \nonumber 
\end{eqnarray} 
and using the same countour integral it can
also shown be that $L(\omega)$ is properly normalized:
$\int_{-\infty}^{\infty}\frac{d\omega}{2\pi}L(\omega)=1$.
The function $ G(z) $ is thus completely free from singularities in
the strip $0 <  Im z<\Omega_N$  and has poles outside the strip at 
$\epsilon-i\gamma$ and $\epsilon+i\gamma+i\Omega_N$ and obeys
$G(z) = G( z + 2 i \Omega_N )  $ as shown in Figure \ref{analyticG}a.  In the limit 
$ \eta \rightarrow 0 $ the function
reduces to $G(z)=\frac{1}{z- ( \epsilon  -  i\gamma ) }$
which is the usual retarded Green's function from a
Lorentzian spectral weight. Evaluating $G(z)$ in the strip
$-\Omega_N<Im z<0$ gives an expression which is analytic in that strip  and
which analogously corresponds to the advanced Green's function. The full
analytic Green's function with periodically repeated branch cuts corresponds to gluing together the two
branches as indicated in Figure \ref{analyticG}b. In the limit
$\eta\rightarrow 0$, $\Omega_N=\pi/\eta\rightarrow\infty$
thus reproducing the standard structure of the analytic Green's
function with a branch cut on the real axis.

\begin{figure} \label{fig:discplot}
\includegraphics[height=5.2in]{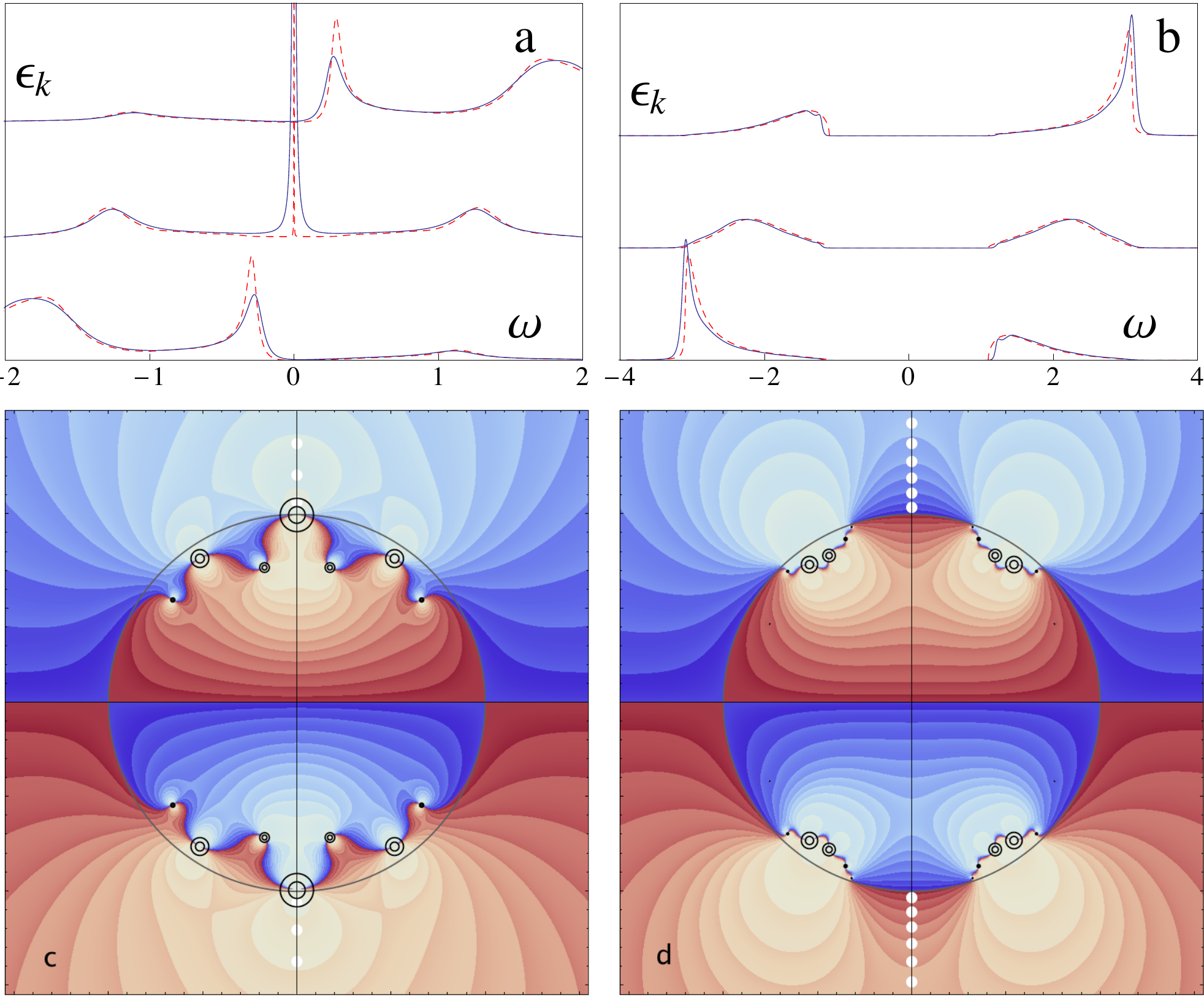}

\caption{\label{spectra}
(Color online) Spectral functions $A(\ek,\omega)$ at $\beta=25$ for metal $U=2$  (a) and insulator
$U=4$ (b) for bare energies $\ek=-1,0,1$. (Dashed curves are $\beta=500$ and $\beta=200$ respectively.) 
The corresponding $ Im( G_{\epsilon_k=0}( z' ) ) $ is shown in (c) and (d) including circles marking the location
of poles.   The poles are located inside
the unit circle in accordance with Eq. \ref{G_transform}. 
Empty dots represent a subset of the  Matsubara
values ($z'(i\omega_n)$). The color scale is light for large
positive and negative values with dark colors near zero.} 
\end{figure}

\begin{figure}
\includegraphics[height=4.8in]{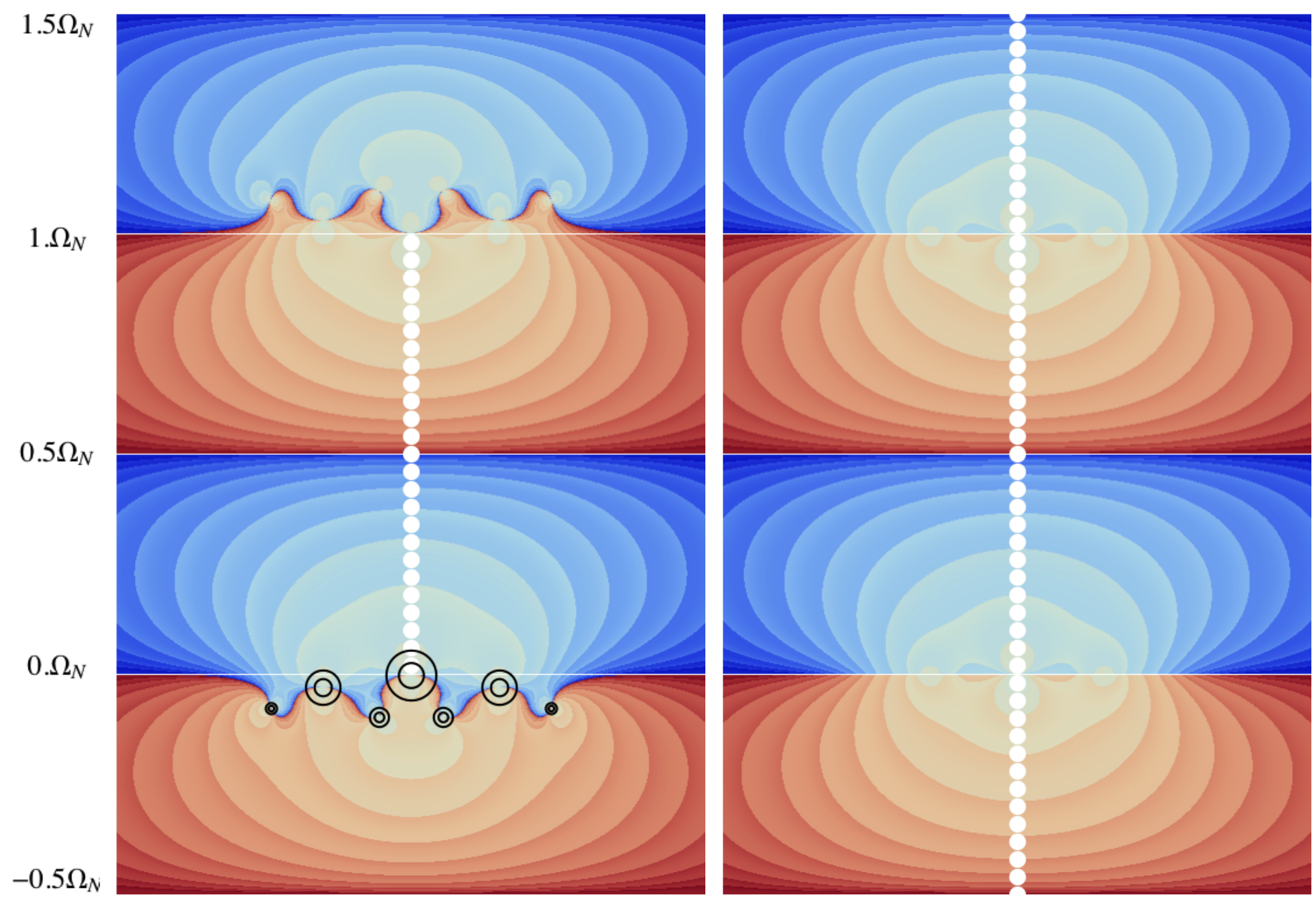}
\caption{\label{analyticG}  (Color online) Analytic structure of the Green's function
  $G_{\ek=0}(z)$. In (a) the branch ($G^R$) which is analytic in
  $0<Im\, z< \Omega_N$. $Im(G)$ is plotted as in Fig. \ref{spectra}c. In (b) the corresponding periodic Green's function with 
branch cuts at integer multiples of $i\Omega_N$. ($Im z$ in units of $\Omega_N=\frac{2\pi N}{\beta}$)}
\end{figure}

Let us now consider a spectral function defined by a set of such 
``periodized Lorentzians'' $L_{\epsilon_\nu,\gamma_\nu}(\omega)$.  For $0 < Im\,z <  \Omega_N$
we find
\begin{eqnarray}
\label{Gofz}
G(z)= \frac{\eta}{2} \sum_\nu
[a_\nu\coth \frac{\eta}{2} (z-\epsilon_\nu+i\gamma_\nu)+\\ \nonumber
a^*_\nu\tanh \frac{\eta}{2} (z-\epsilon_\nu-i\gamma_\nu)]\,,
\end{eqnarray}
where we have allowed for a complex prefactor $a_\nu$ that must
satisfy $\frac{1}{2}\sum_\nu(a_\nu+a^*_\nu)=1$ to conserve total
spectral weight.
Given a set of values of $ G_n  \equiv G_k( i \omega_n ) $
we would like to extract the values of $a_\nu$, $ e_\nu $ and $ \gamma_\nu $ 
that solves Eq. \ref{Gofz}. The latter can be cast into the form of a 
rational function containing only simple poles (as opposed to periodically repeated) 
by means of the conformal mapping 
\begin{equation}
z'=\coth(\frac{\eta}{2}z-i\frac{\pi}{4})
\end{equation}
which gives
\begin{equation}
G(z')= \frac{\eta}{2} \sum_{\nu=1}^{n} \left(a_\nu\frac{1-p_\nu z'}{z'-p_\nu}+a^*_\nu\frac{1-p^*_\nu z'}{z'-p^*_\nu}\right)\,,
\label{G_transform}
\end{equation}
with $p_\nu=\coth( \frac{\eta}{2} (\epsilon_\nu-i\gamma_\nu)-i\frac{\pi}{4})$.
The transformation maps the strip $0 \le Im z< 2 \Omega_N$ to the entire complex plane.
The strip $ \Omega_N  \le  Im \ z <   2\Omega_N $ that contains the singularities
is mapped to the interior of the unit circle and
the strip $ 0 < Im\, z < \Omega_N $ that is free from singularities is
outside, as exemplified in Figure \ref{spectra}c and d.  The points $ z =  \pm \infty $ on the real axis map to
$ z' = \pm 1 $ and the point $  z =  i\omega_{(N-1)/2} $
maps to $ z' =  \infty $.  

The function $ G( z') $ should obey a number of properties.
Since Eq. \ref{G_transform} can be written as a rational function
$ G( z') = const + P(z')/Q(z') $
where  $ P $ is an $ M-1 $ degree and $ Q $ is an $ M'th $ degree polynomial
of $ z' $ and $ const = - \frac{\eta}{2}  \sum ( a_\nu p_\nu  + a_\nu^* p_\nu^* ) $,
we can identify $ p_\nu $ with the roots
of $ Q $ and $ \frac{\eta}{2}( 1 - p_\nu^2) a_\nu $ as the residues of
$ P/Q$. Taking $ N $ odd gives
a precise boundary condition $ G(z'=\infty) = const = G_{(N-1)/2} $.
We also require $G(z'=1)=-G(z'=-1)= \frac{\eta}{2} \sum_\nu(a_\nu+a^*_\nu)=\eta$
to obtain the proper normalization.  A crucial observation is that
Eq. \ref{eq:newnewdyson} preserves this condition independent 
of $ \Sigmabar $ and demonstrates that total spectral 
weight is preserved by the periodized Dyson equation.  
By straightforward calculation from  Eq. \ref{GofA} we also conclude that
$ G'(z=1) = G'(z=-1) = 0 $ which further constrains the analytic continuation.
We thus find $ P(z') $ and $ Q(z') $ by fitting
$ G(z') - const $ to $ (M-1)/M $  Pad\'e form, yielding $ M =  (N+3)/2 $.
In addition we have the symmetry $G(i\omega_{N-1-n})=G^*(i\omega_n)$
resulting in $ M/2 $ independent poles. Thus $ n = (N+3)/4 $ in Eq. \ref{G_transform}.
Having identified the parameters $ a_\nu $ and $ p_\nu $ in the fit, the spectral function can
be evaluated through $A(\omega)=-2 Im G(\omega)$ using Eq
\ref{Gofz}. 
For every value of $ N $ we can also obtain a properly normalized
spectral weight using $ A(\omega)  = 2\sum_{\nu} \frac{ Re a_{\nu}\gamma_{\nu}-Im a_{\nu} (\omega-\epsilon_\nu)}{ ( \omega-e_{\nu} )^2 + \gamma_{\nu}^2 } $
resulting in  Green's function with the proper discontinuity
$G_k(\tau=0^-)-G_k(\tau=0^+)=\int_{-\infty}^{\infty}\frac{d\omega}{2\pi}A(k,\omega)=1$.

We tested our method on the Dynamical mean field theory (DMFT)
method using the IPT approximation with the half filled paramagnetic
Hubbard model.\cite{George_Kotliar92,George_review,Potthoff,Hugo} 
We found that the
convergence was significantly  
enhanced by evaluating the dynamic part of the self energy as 
$\tanh \etatwo \Sigma(i\omega_n)= \etatwo U^2 {\cal F} \left( G^3_{0,imp}(\tau_j) \right) $
where Eq. \ref{eq:coth} has been used.
This expression ensures that $Im \Sigma$ lies within the
appropriate bounds and by expanding $\tanh$ in powers of $\Sigma$
it is equivalent to the standard expression to order $U^2$. The
detailed implications of this procedure should be explored further.

Figure \ref{spectra} shows the spectral function for
varying bare energy $A(\epsilon_k,\omega)$ at $\beta=25$ ($T=W/50$),
in the metallic, $U=2$ (with $N=25$), and insulating, $U=4$ ($N=45$) phases using the
standard semicircular bare density of states\cite{Hugo}
and 40  significant digits to compute the rational function. This gives
7 and 12 independent complex poles respectively.
The location of poles are indicated in Fig. \ref{spectra} for $\epsilon_k=0$.
Since $\Sigma$ is $k$-independent we can extract it with a single
Pad\'e fit and use this to generate the spectral function for any
$\epsilon_k$. 
We have found that the number of frequencies needed
is roughly $  N \gtrsim \beta $  for convergence
of the IPT recursion algorithm 
and as further test we have used the method at low temperature 
$ \beta \geq 500 $ and a number of significant digits roughly also equal 
to $ N $.

As a second test of the analytic continuation method, we considered
a noninteracting single impurity Anderson model whose spectral function 
consists of a sharp resonance as well as a continuum. These
features have been found to be difficult to reproduce in detail
using ordinary Pad\'e methods.\cite{schon} We used Eq. \ref{GofA}  to
compute $ G_{imp}(i\omega_n) $ from the exact $ A_{imp}(w) $ using
$N=61$, $\beta=25$, 
and 60 significant figures. Using this we find 16 independent
complex poles that to the eye reproduces the exact spectral function.
It has an integrated rms deviation over total weight (excluding the
resonance) of $4\cdot 10^{-3}$.
The resonance has a width of $ 10^{-8} $ and a 
normalization which is within $ 10^{-5} $ of the exact value. A major
reason for our success is the high precision for the Pad\'e fit rather 
than a large number of poles.  Fitting a greater
number of less accurate data points does not yield comparable
accuracy.  This motivates using our method to compute a Green's function 
to extremely high precision in a relatively small dimensional space and
to use the combination of conformal transformation and Pad\'e fit to
infer the analytic continuation to the real axis.

We have presented a method to work with discrete-time Matsubara
Green's functions.  In spite of working in a finite dimensional space
the method yields an analytic continuation that obeys the boundary
condition at $ \tau = 0$, a problem that has plagued other previous
discretization schemes. Within this small space,
numerical calculations can be done to the extremely high accuracy
that is necessary to numerically obtain a meaningful analytic
continuations from imaginary to the real time axis. The method
should have wide applicability to all problems requiring numerical
evaluation of Matsubara Green's function in condensed matter physics
nuclear physics and quantum field theory.


\begin{thebibliography}{10}
%

\bibitem{ref:abrikosov} A.A. Abriksov, L.P. Gorkov, I.E. Dzyaloshinskii, {\em Quantum field theoretical methods in many body physics.}, (2ed, Pergamon, 1965).
\bibitem{N_O} J.W. Negele, and H. Orland, ``Quantum Many-Particle Systems'', Addison-Wesley, 1988.
\bibitem{vidberg}H. J. Vidberg and J. W. Serene, {\em  J. Low Temp. Phys. }{\bf 29 }, 179 (1977).
\bibitem{Jarrell} R.N. Silver, 
J.E. Gubernatis, 
D.S. Sivia,  
and M. Jarrell, Phys. Rev. Lett. {\bf 65}, 496 (1990).
\bibitem{numexact} By numerically exact we mean a convergent solution 
computable to any number of significant figures with a realistic  amount of computer time. Calculations
were done  using arbitrary precision arithmetic in {\em Mathematica}. We have had no
trouble performing calculations with several hundred significant digits in order to resolve 
a large number of poles in the Greens function.
\bibitem{note1} It should be noted for {\em any} calculation for regularly spaced imaginary time coordinates, 
be it quantum Monte Carlo or perturbative Greens function calculations, the natural expansion 
is a Greens function of the form Eq. \ref{eq:coth} that fits all the ``Matsubara'' data rather 
than a direct fit of the continuum form $ 1/( i \omega_n - e_k ) $.
%
\bibitem{LW} J.M. Luttinger, and J.C. Ward, Phys. Rev. {\bf 118}, 1417 (1960). 
%
\bibitem{Baym_Kadanoff} G. Baym,
{\it Progress in nonequilibrium Green's functions},
Proceedings of the conference,  ``Kadanoff-Baym Equations Progress and Perspectives for Many-body Physics'',  Rostock Germany, 20-24 September,  1999.
\bibitem{George_Kotliar92} A. Georges, and G. Kotliar, Phys. Rev B {\bf 45}, 6479 (1992).
%
\bibitem{George_review} A. Georges, G. Kotliar, W. Krauth, M. Rozenberg, Rev. Mod. Phys. {\bf 68}, 1996.
\bibitem{Potthoff} M. Potthoff, Adv. in Solid State Physics, {\bf 45,},  135 (2006); Eur. Phys. Journal B, {\bf 36}, 2003.
%
\bibitem{Hugo} H.U.R. Strand, A. Sabashvili, M. Granath, B. Hellsing, S. \"Ostlund
{\em Phys. Rev. B }{ \bf 83 }, 205136 (2011).
\bibitem{schon}  C Karrasch,R Hedden, R Peters, Th Pruschke,K Schonhammer and V Meden, 
{\em J. Phys.: Condens. Matter }{\bf 20 } 345205 (2008); C. Karrasch, V. Meden, and K. Sch\"{o}nhammer,
{\em Phys. Rev. B }{\bf 82 }, 125114 (2010).
\end{thebibliography}
\end{document}